Mind the Dark: A Gamified Exploration of Deceptive Design Awareness for Children in the Digital Age

This paper addresses the critical issue of deceptive design elements prevalent in technology, and their potential impact on children. Recent research highlights the impact of dark patterns on adults and adolescents, while studies involving children are scarce. In an era where children wield greater independence with digital devices, their vulnerability to dark patterns amplifies without early education. Our findings show a significant positive impact of dark pattern education on children's awareness, revealing that heightened awareness considerably alters children's navigation of social media, video games, and streaming platforms. To this end, we developed a gamified application aimed at instructing children on identifying and responding to various dark patterns. Our evaluation results emphasize the critical role of early education in empowering children to recognize and counter deceptive design, thereby cultivating a digitally literate generation capable of making informed choices in the complex landscape of digital technology.

**Authors:** Noverah Khan, Hira Eiraj Daud, Suleman Shahid; from the Lahore University of Management Sciences

## 1 INTRODUCTION

In light of the current rate of technological advancements, digitalization has found a home amongst almost all age groups in society. This directly follows an increase in the frequency of websites and mobile applications tailored to each age group, featuring "persuasive approaches to influence human behavior and decision-making" [1]. While aligning with the psychological and mental model of the user is a testament to good design of user interfaces, specific malicious tactics to deceive users have come to abuse the power of design patterns. A Dark Pattern is an interface that has been designed to deceive users into performing actions that they did not mean to [2] or that goes against their best interests [3] or that benefits the designer at the expense of the user's welfare [4]. These tricks can be subtle or overt and children tend to fall prey to these manipulative patterns. Lack of awareness in detecting and reacting to dark design patterns entangles children into a vicious cycle of manipulation and implicit coercion and thus, it is critical to teach children how to detect and deal with dark UI.

While existing literature creates a framework for categorizing dark patterns and how they might affect adolescents in general, our research targets a specific age group. According to Jean Piaget, a child's concrete operational stage of cognitive development begins from approximately the ages of 7 to 11 [5]. It is at this stage that children begin to understand logic and reasoning, as well as when they gain more technological autonomy [6]. According to the Information Commissioner's Office's (ICO) code for age appropriate design, children go through 5 stages of development [7]. Ages 6-9 are termed the "core primary school years" and it is at this stage where children are more likely to use devices independently. It is also at this age where they are likely to develop a basic understanding of online safety and digital environment which they might learn at school. Ages 10-12 which are defined as the "transition years" is the key age at which children's online activity is likely to change. At this age, they are likely to own their own personal device such as a phone and they begin to develop a better understanding of the online world, however, they are still unlikely to be aware of the full extent of the use of their personal data [7].

To this end, we identify the ages of 8 to 12 as the core age group for our research concentration. In this early stage of their technological understanding, we aim to develop children's mental models so they can prevent uninformed and accidental decisions while using technology, in conjunction to improving their tech literacy and self-reliance.

## 2 RELATED WORKS

### 2.1 Impact of Dark Patterns on Children

The effects of dark patterns may be "disproportionate" to children. [4]. Dark patterns are hazardous for adults and children, notably when they use social media or play online games [4, 8]. Almost every pre-teen has access to the internet and uses mobile and gaming devices, allowing for constant engagement [6, 9], transforming how the industry views them. Compared to what adults endure, adolescents are more vulnerable to the dangers of social media [10]. Lupton and Williamson describe the "dataviellance" of children and how digitized strategies of covert surveillance like continual monitoring of online interactions has given internet companies continuous access to children's data to exploit for commercial purposes [11]. Using this data, companies have been creating more curated online experiences for children, using algorithms that track and predict their preferences making it addictive to use [12]. Addictive design can have a negative impact on children's mental health, strain relationships, and increase exposure to cyber bullying [12]. Additionally, excessive technological use presents a huge opportunity cost in children's time, where they feel compelled to continue using their devices even when they need or want to do something else [12]. As the media continues to commercialize young children, it demands an investigation into the gap in the research on the relationship between dark design and privacy and monetization of mobile apps for children [8]. Educating adolescents about dark designs is vital so that they not only develop into digital-literate users but also resist the influence of manipulative designs [10].

### 2.2 Dark Patterns in Children's Apps

Dark patterns have become common in mobile games and apps aimed at children [13]. Zagal et al. categorize dark patterns found in games as either temporal, monetary, or social [14]. Temporal dark patterns such as "play by appointment," restrict the user's ability to play only when the game allows it. Games like "Pokemon-GO" and "Animal Crossing" pressurize the user to play at inconvenient times as an obligation rather than playing at their leisure [14]. Monetary dark patterns like "pay to skip" extract money from players by allowing them to skip levels or challenges by paying money. Zagal et al. discuss how apps such as "Clash of Clans" and "Angry Birds" let players pay real money to avoid the wait imposed after building something or purchase power ups that can be used to pass a level automatically [14]. Social dark patterns endanger a player's social standing and relationships. The "Social Pyramid Scheme" only allows the player to proceed through the game if they invite friends and other people to play. For example, "Farmville" requires having other players as neighbors, resulting in persuading real-life friends to join [14].

Video streaming platforms profit from increased active user participation. The "infinite scroll", Netflix homepage trailers, hidden signouts, and the YouTube autoplay feature all undermine the individual's autonomy by feeding them with more content in an effort to keep them on the app. Roffarello et al. describes how features such as autoplay aim to diminish the user's cognitive or physical ability [3]. Most of these use a "variable-rewards schedule approach" to make sure users are active and hooked [4].

Social media sites employ what Lukoff et al. [15] define as "attention-capture" dark patterns. In order to ensure that users spend as much of their time as possible on social media, their interfaces are designed in a way to make it difficult to leave. Milder and Savino discovered that Facebook has changed the position of their log out button multiple times making it harder for users to exit the application [16]. Roffarello et al. mention how the "pull-to-refresh strategy" misleads the user to encourage continued usage of the platform by, for example, sending new messages [3]. Additionally, Janmohamed details how many sites use machine-learning powered recommendation systems [17]. They recommend groups to join or accounts to follow and they learn from user's responses to these recommendations in order to curate their social media experience further [17].

### 2.3 Community Impact on Children's Interaction with Dark Patterns

Linebarger et al. discovered that parenting quality could impact the relation between "media use and child development" such that "inappropriate content and inconsistent parenting" negatively affected the executive functions of low-income preschool-aged children. In contrast, responsive parenting focused on educational content brought about greater benefits for a child's development [18]. Thus parental involvement is likely to positively impact children's use of technology. Munzer et al. suggest the regulation of children's consumption of digital mobile media may have an impact on self regulation in children, characterized by measures of self-control along inhibitory, emotional, and attentional axes that reflect in our emotional and behavioral responses to challenging situations and includes, for instance, the ability of a child to "delay gratification: postponing immediately available smaller rewards while waiting for larger rewards" [19]. This situation could be mapped to children being exposed to disruptive in-app advertisements in a malicious attempt to prolong gameplay [20]. Therefore, difficulties in self-regulation could imply greater vulnerability to falling for dark design traps. This proves the necessity of teaching children about internet safety in a monitored environment.

There is merit in receiving an education about deceptive UI and internet safety in a collaborative learning environment and schools function as a critical place to receive online education. According to Data Matters [21], teachers play a crucial role in supporting children online, with 70% of children turning towards teachers for information about staying safe online. Additionally, online safety is part of school curriculum, and schools often make up for any deficiencies in online safety education and oversight that children might experience from their parents. Albert Bandura's theory of social learning emphasizes the importance of observational learning, where individuals acquire knowledge, skills, and beliefs by watching the actions of others, leading to the modeling and adoption of observed behaviors [22]. Social learning improves knowledge retention [23]. Additionally, if dark patterns were brought up as a conversation within their schools, children would be more likely to learn and implement the knowledge that they learn. Collaborative learning techniques enhance critical thinking amongst students [24]. This might also build a sense of community around dark patterns, emphasizing that it is not just the individual's job to deal with them but a community effort to maintain a standard of education and help each other safeguard themselves on the internet.

## 2.4 Governmental Approach

The dangers of persuasive design elements have not been lost on governmental organizations, with many policy approaches and class action suits filed against corporations for consumer protections. The OECD published a report on their roundtable conference on dark patterns, where they acknowledge that children are more likely to be vulnerable to dark patterns [25]. The Information Commissioner's Office developed an age appropriate design code of practice for online services [7]. The code encompasses the design techniques that information society services (ISS) must follow based on the age bracket of the children that access their websites in the UK. Similarly, in their guidelines for dark patterns on social media interfaces, the European Data Protection Board highlights the increased risk of children to these deceptive design interfaces [26]. They state that children may be less aware of the risks associated with the processing of their personal data, which is why they mandate that companies make it clear to children what data they are consenting to being processed. They also state that children are more likely to be influenced by "emotional steering", i.e design decisions which influences users emotional state in a way that is likely to lead them to act against their data protection interest. Thus, social media platforms should ensure that the language used to address children about information should be easily understood by them.

The US Federal Trade Commission (FTC) has taken action against Amazon, Apple, and Google for placing hidden charges in free apps meant for children resulting in Apple and Google refunding their customers $32.5 million and $19 million respectively [27,28]. Similarly, the children's online learning program ABCmouse automatically renewed subscriptions and charged users without their consent, which resulted in them paying a $10 million fine [29] and Epic Games and Fortnite were ordered to pay half a billion dollars in penalties and fines [30] for violating children's privacy laws and tricking users into making unwanted payments.

While there exists a framework for companies in these countries to follow to prevent consumer deception, that has not stopped them from continuing dark pattern practices, as evidenced by the multiple lawsuits taken against the same companies. So while the onus on consumer protection falls on government entities, policy takes time to design and implement, and oftentimes, it fails to be enough. Since corporations constantly invent new ways to undermine consumer protection laws, inevitably, consumers must take it upon themselves to be educated on deceptive design in order to avoid it. Although, educating children about dark patterns is not enough to fully protect themselves from it, education spreads awareness, which is necessary for children to be able to identify deceptive design to begin with. Thus, digital literacy is not meant to replace regulatory efforts but instead to complement them.

While previous research has identified the presence of deceptive design in children's apps, labeling and categorizing the types of dark patterns present, existing literature has yet to explore how young children directly understand, engage with, and deal with these dark patterns. This gap in our understanding necessitates an investigation into how children interact with deceptive design especially at an age where they are now experiencing more technological autonomy. Based on the aforementioned review, we identify the specific questions we seek to answer below:

> **RQ1:** Are children aged 8 to 12 aware of dark patterns? How much they know about them, can they recognise them, and how do they react to them?
> **RQ2:** What effect does receiving an education on dark patterns have on preparing children to identify deceptive design patterns in an effort to avoid them?
> **RQ3:** How can we design an effective educational intervention for children that could be easily implemented in their daily routines?

## 3 METHODOLOGY

We employed a mixed-methods approach consisting of surveys and focus group activities. As identified above, our target audience spans children aged 8-12, which corresponds to primary and middle school. We conducted 6 in-class focus groups with 147 children from grades 3 to 7 in a local private school in Lahore, Pakistan. Firstly, we noted down the demographic information of each group. We then presented them with scenarios about dark patterns and recorded their responses to them. Next, we gave a short lecture explaining what dark patterns were, the different ways they could be present within applications and games, and how they could potentially harm users. After our lecture, we showed the children the same scenarios and recorded their responses to see if there was a change in the action they took. Lastly, we opened the floor for discussion, letting children share their personal experiences with deceptive design and any opinions they had.

## 4 USER RESEARCH

### 4.1 Demographic Results

The following table shows the distribution of ages across the grades.

Table 1. Distribution of ages across grades 3-7

| Age | Grade 3 | Grade 4 | Grade 5 | Grade 6 | Grade 7 | Total |
|---|---|---|---|---|---|---|
| 7 | 3 | 2 | 0 | 0 | 0 | 5 |
| 8 | 13 | 4 | 0 | 0 | 0 | 17 |
| 9 | 12 | 17 | 6 | 0 | 0 | 35 |
| 10 | 0 | 3 | 13 | 4 | 0 | 20 |
| 11 | 0 | 1 | 1 | 30 | 9 | 41 |
| 12 | 0 | 0 | 1 | 12 | 12 | 25 |
| 13 | 0 | 0 | 0 | 2 | 2 | 4 |
| Total | 28 | 27 | 21 | 48 | 23 | 147 |

The scatter plot in Figure 1 shows students' daily average screetime and on average, children had a screentime of greater than 5 hours. Students were then asked what types of devices they used, the results of which are depicted in the bar chart below. Most children used smartphones.

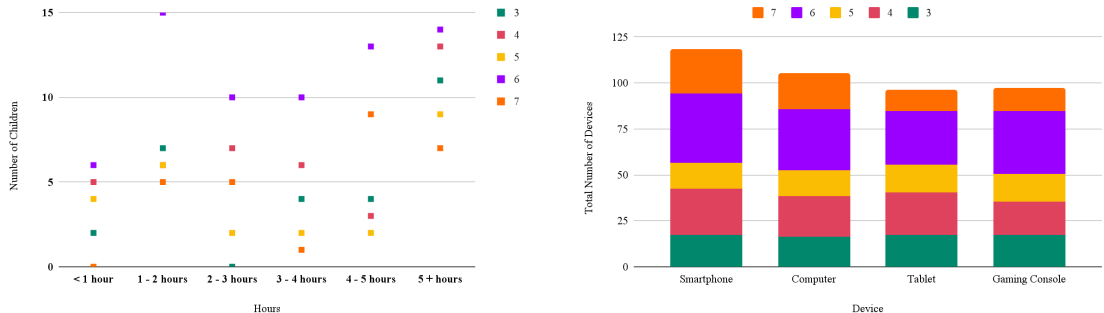

Fig. 1. The screentime of children from each grade (scatterplot); the types of devices used by children (bar graph)

Next, we asked children if their parents monitored their screen time and content on their devices. Our findings revealed that for most students, their parents did monitor their screen time and the content they consumed, however there was a drop in the monitoring of grade 7 with only 25% of kids' parents keeping their digital activities in check.

In an extended focus group conducted with 16 students of grade 5, children were asked to write down their most commonly used applications, the top 3 being YouTube (62%), Snapchat (37%) and Roblox (31%).

### 4.2 Lecture and Test

To gauge how children would react when faced with dark patterns we performed a test comprising three scenarios taken from the game "Dragon City" [31], with each scenario based on a different category of dark patterns as identified by Gray et. al [32]. Our first scenario was based on the "confirm-shaming" tactic. Children were shown an emotionally manipulative pop up from the game which asked them to free a caged dragon by purchasing gems with real money. The dragon was crying, in order to guilt trip users into making the purchase. Our second scenario, which was based on the 'fake scarcity' tactic, showed a pop-up offering a 92% discount with a one-time offer and a time limit to purchase a 'legendary' dragon. Our third scenario was a follow-up to the second scenario and was based on the 'nagging' tactic. In case the user decided not to make a purchase of gems for the 'legendary offer', the game showed a pop-up with the mascot saying 'The offer is only available now, are you sure you want to miss it?'. The pop-up involved emotional manipulation depicting teary-eyed baby dragons with broken-hearted emojis and the 'trick wording' tactic with the following two course-of-action options - (1) Go Back and (2) Lose Offer. The 'go back' option confused users into thinking it would take them to the main page, whereas it actually simply took them back to the legendary offer pop-up.

After the test, we delivered our lecture, using material different from the test. We discussed different types of dark patterns with examples, including nagging, fake scarcity, hard to cancel, trick wording, confirmshaming, and emotional manipulation. After the lecture we performed the test again. The table below shows the responses of students for the 3 scenarios, with each scenario having 3 kinds of responses - (1) "Purchase" or "Go Back" meaning that they decided to spend money to buy gems, (2) "Exit" or "Lose Offer" meaning they reject the offer and move on in the game, and (3) "Unsure" meaning they were uncertain about how to proceed.

Table 2. Student responses for the 3 scenarios based on dark patterns, with each response recorded before and after the

| S No. | Response | Before Lecture | | | | | After Lecture | | | | |
|---|---|---|---|---|---|---|---|---|---|---|---|
| | | G3 | G4 | G5 | G6 | G7 | G3 | G4 | G5 | G6 | G7 |
| 1 | Purchase | 9 | 16 | 7 | 17 | 5 | 1 | 0 | 1 | 4 | 0 |
| | Exit | 10 | 11 | 13 | 31 | 18 | 16 | 27 | 36 | 44 | 23 |
| | Unsure | 9 | 0 | 1 | 0 | 0 | 11 | 0 | 0 | 0 | 0 |
| 2 | Purchase | 10 | 13 | 2 | 5 | 7 | 0 | 0 | 1 | 2 | 4 |
| | Exit | 11 | 13 | 13 | 43 | 14 | 28 | 27 | 20 | 46 | 19 |
| | Unsure | 7 | 1 | 6 | 0 | 2 | 0 | 0 | 0 | 0 | 0 |
| 3 | Go Back | 10 | 10 | 5 | 6 | 1 | 0 | 0 | 1 | 2 | 0 |
| | Lose Offer | 11 | 10 | 17 | 42 | 17 | 28 | 27 | 20 | 46 | 23 |
| | Unsure | 7 | 7 | 1 | 0 | 5 | 0 | 0 | 0 | 0 | 0 |

lecture. *S No: Scenario Number *G: Grade

Table 2 displays a clear drop in the number of students, across all the grades, who decided to make a purchase i.e. spend actual money for all 3 scenarios after the lecture. Moreover, there's a rise in the number of students who decided to exit i.e. reject the offer after the lecture. Finally, there was an overall decrease in the number of students post-lecture who were unsure of their responses initially. Thus, it can be concluded from the activity that the information session on dark patterns increased children's awareness about it while reducing uncertainty.

### 4.3 **Findings**

Through children's responses during the activity, we noted that while they did not know any formal terminology to define dark patterns, they could identify some of the interfaces that were designed to trick them. Many children, especially those from middle school, had knowledge of data and phishing scams and could determine the underlying motivation of the dark patterns that we showed them. Many cited personal experiences of their relatives being scammed by unknown callers. From our observations, children across all ages had a general understanding of what experiences could be classified as "dangerous" or "wrong", however, they could not further categorize any instances beyond the binary terms of good or bad. Children across all the classes consistently used the word "hacker" to describe anyone with whom they had an unpleasant encounter with on the internet. And even though all the instances mentioned by the children were concerning, they mostly reflected criminal behavior rather than the legal-but-problematic practice of creating manipulative interface elements, which indicated that the children were generally able to identify 'extreme' cases of deceptive design but would overlook the small design details or cleverly-included design elements e.g., hard-to-find or smaller-than-usual-sized close buttons, false scarcity text etc. Another important observation we made was that during the lecture and activity, children began discussing their experiences amongst themselves and the group discussion encouraged more children to speak out about their personal experiences after hearing their peers talk about theirs.

During the activity, students raised concerns and shared their experiences with regards to cyber malfeasance around the following two themes:

*4.3.1 Encounters with Strangers.* Most of the children admitted to having unwanted and uncomfortable encounters with strangers in video games and on social media platforms, which involved inappropriate content or unsecure URLs. They mentioned having conversations with their parents about strangers (in real life) and being instructed to stay away from them. Thus, the majority stated that they had the same reaction to these 'strangers on the internet', with some students 'running away' (not further engaging in the online conversation with the strangers) while others going to their parents for assistance.

Many children's experiences with data phisers was in the game "Roblox". The game allows users to speak directly with other players. One student spoke of a negative interaction in the game:

"This person kept trying to talk to me and kept following me around in the game..I ran away" [P1, G5].

Another student recounted a similar experience:

"I was playing Roblox, and this stranger kept following me and then he sent me a request on Facebook and then he started sending me weird messages and links and my account got hacked" [P2, G5].

Students also discussed their experiences with dark patterns on social media platforms like Facebook and Snapchat, involving strangers and fake profiles attempting to hack their social media accounts. A student shared his experience about accidentally accepting a stranger's social media request and consequently receiving inappropriate images and prying questions about personal information:

"I accidentally added a stranger on Snapchat, and they started asking me personal questions and they sent me bad pictures so I told my dad and he reported them" [P4, G5].

Another student also recalled a similar experience with a fake social media profile trying to trick her into clicking on a scam link in order to hack her account:

"A fake account on Facebook messaged me saying 'your data got deleted, please give your info to recover it'. I think it wanted to hack my account or get my pictures" [P5, G6].

*4.3.2 Deceptive Design Experiences.* Besides having uncomfortable encounters and conversations with strangers through video games, students talked about their experiences involving online purchases and monetary scams within the games they most frequently played. The games would repeatedly show limited-time offers, rare objects purchasable through actual money and free game-currency (gems or coins) deals which would show the actual cost after users clicked on it. A student shared his experience with the hidden costs dark pattern on Roblox:

> "There was a free Roblox pop-up but when I clicked on it, it said $500" [P3, G3].

Another student shared her experience with the hidden cost dark pattern on her browser:

> "I was randomly surfing the internet and on one website I came across this pop-up which said 'free money' but when I clicked on it, it asked for my personal information, so I closed the website" [P4, G6].

A student also discussed her experience with the pay to win dark pattern:

> "I remember FIFA constantly asking me to pay to get more points to win the game." [P5, G7].

Thus, some of the monetary dark patterns we identified through the experiences shared by the students included hidden costs, bait and switch, artificial scarcity and pay to win.

Students also spoke about their encounters with dark interface patterns on video streaming platforms like YouTube, Netflix and HBO Max, which involved deceptive elements carefully designed to make the users stay on the platforms for longer or avail special offers. Some of the examples mentioned by the students included autoplay, emotional manipulation and nagging. A student shared his uncomfortable experience with YouTube autoplay, which according to him, started showing inappropriate videos from its suggested videos section:

> "My YouTube started showing bad videos on its own through autoplay" [P6, G5].

Another student talked about her experience with 'nagging' on YouTube:

> "YouTube Premium pop-ups keep nagging me with the no-add offer and then YouTube shows so many long ads that I sometimes want to go for it [the YouTube Premium Subscription] but then I remember that it's my mom's device and also I don't have a card obviously 'cause I'm a kid" [P7, G7].

Other students discussed how Netflix and HBO Max used the emotional manipulation dark pattern, sending emails on their parents' devices with 'sad emojis', asking them to not cancel their subscription or to renew their subscription again.

## 5 DESIGN SOLUTION

### 5.1 Design Objectives and Process

Our research highlighted that while children were able to identify elementary manipulative design, they possessed no knowledge of the formal terminology of dark patterns nor why anyone would want to use them. Furthermore, our research revealed the magnitude of dark patterns used in the apps that children used every day. Our discovery necessitated an intervention aimed to educate children about the existence and impact of dark patterns in the online realm. Our objective for this intervention was to empower children against manipulative design by equipping them with the tools to effectively identify and resist deceptive UI by exposing them to examples of dark patterns in familiar applications. Additionally we aim to improve children's digital literacy while providing practical guidelines for developing healthy online habits.

As evidenced by our literature review, children whose online experiences have a degree of parental involvement are better at practicing self regulation and schools function as an effective site to receive online safety education. Thus we knew that our intervention needed to be one that could be introduced in schools. According to Carla Fisher, in order to design games to be used at school, we should leverage insights from teachers who may be able to familiarize us with school infrastructure [33].

Thus, we decided to embark on a collaborative design process with 5 middle school teachers, in order to leverage their expertise into designing an intervention best suited for our target age group. Teachers prioritized engaging children over anything else, as any attempt to educate them would be futile if they didn't want to learn in the first place. To do this, they pointed out that children like first-person games such as Roblox or Minecraft since it gives them a sense of control. Additionally, personalisation tactics such as being able to design their own avatar, participation certificates, and "cute" rewards would be a good way to incentivise children to use our intervention. In order to make our intervention age appropriate, they recommended utilizing lots of colors and using animations and graphics in lieu of text, since children avoid reading large chunks of text. Teachers advised us to take inspiration from the application "Taleemabad" [34], which is an educational app that they used within classes as well. They stated that the lecture to game to test based flow of "Taleemabad" keeps children engaged while they learnt new content and constantly reinforced the new information that they had just learned. In order to improve their retention of new knowledge, teacher's suggested that we add constant reinforcement in the form of revision or testing every time a new concept is brought up. The structure of knowledge reinforcement that teachers suggested is inline with "HTML Heroes", an internet safety programme co-funded by the European Union with the same target age group as our intervention [35]. HTML heroes' lecture to quiz based model, supplemented by engaging animations has proved to be successful with Webwise reporting a 25% increase in the use of its resources when the program launched [36].

During our user research phase, our initial target age range comprised both primary and middle school children. However, at the time of co-design, teachers stated that there is a gap between the level of understanding of vocabulary, engagement with animations, and ability to complete challenging games of primary and middle school children. Thus, they pointed out that it would be more effective to design a singular intervention more focused towards either primary or middle school children and hence suggested that we revise our target age range. This is similar to how HTML Heroes has two versions of its game; one for 1st and 2nd grade and one for 3rd and 4th grade [35]. Based on the findings of our user research, we discovered that older children were more adept at identifying manipulative design tactics than younger children. Thus we decided to limit our target age group to just primary school children aged 8 to 10.

Teachers pointed out that education is less engaging and impactful when it occurs in isolation. Which supplements our findings from our literature review. Thus, they recommended that our final solution be one that could be practiced on a larger scale, in a place such as a classroom. In our primary research we discovered that children were taught basic internet safety skills during their computer classes. In agreement with the teachers, we decided that it would be the ideal environment to seamlessly incorporate our intervention.

5.2 **Final Design**

Our final design took the form of a web application tailored specifically for computer classes. The web application would function as a platform for interactive and collaborative learning, allowing the instructor and students to explore and understand dark patterns. Based on the recommendation of teachers, we developed a story-telling based educational intervention. To make the intervention interactive, we incorporated mini-games and quizzes to reinforce the knowledge gained throughout the experience. These interactive elements would test children's comprehension while being a fun way to reinforce key online safety concepts.

The game comprises a virtual world where children can progress through different levels, encountering various scenarios that simulated real-life online situations. At each level, they would face challenges relating to dark patterns. Their ability to recognize and respond to these manipulative techniques would determine their level of progress in the game. The game consists of 4 modules with each one targeting the most commonly experienced dark patterns that children encountered during our user research. These modules were designed to comprehensively understand various dark patterns and their implications in different online contexts. By breaking down the intervention into these focused modules, users can engage with specific aspects of dark patterns. The user would have to successfully complete one module before being able to move onto the next.

Each module follows a three-tier approach, as recommended by the teachers. The first tier is an interactive tutorial which consists of lesson plans that provide examples of and guide children through the concepts of a specific dark pattern. After completing the tutorials, users can test their understanding via a "Test Yourself" section, which allows for self-assessment and reinforcement of the learned material. The second tier is resources which serves as a comprehensive repository of additional resources to assist teachers and students in researching the dark pattern. The last tier is quizzes which provide students with an interactive game-like experience in which they could consolidate their understanding of the newly learned concepts.

The lecture slides used in our primary research included screenshots from the game "Dragon City", which many students stated that they had played before. Our lecture slides were designed using a darker color palette with little text and more colorful graphics. Children were kept engaged during our discussion and many expressed that the visualization of the lecture helped them retain knowledge better. Thus, the visual design of our final intervention was inspired by a "dark and fantastical" theme. This included the use of dark colors, as well as illustrations of fantastical animals, such as dragons, in order to capture children's attention. During our research, we discovered that this theme piqued the children's interest, with dragons being especially popular. Purple and black were the primary colors used, with green, yellow, and red accents.

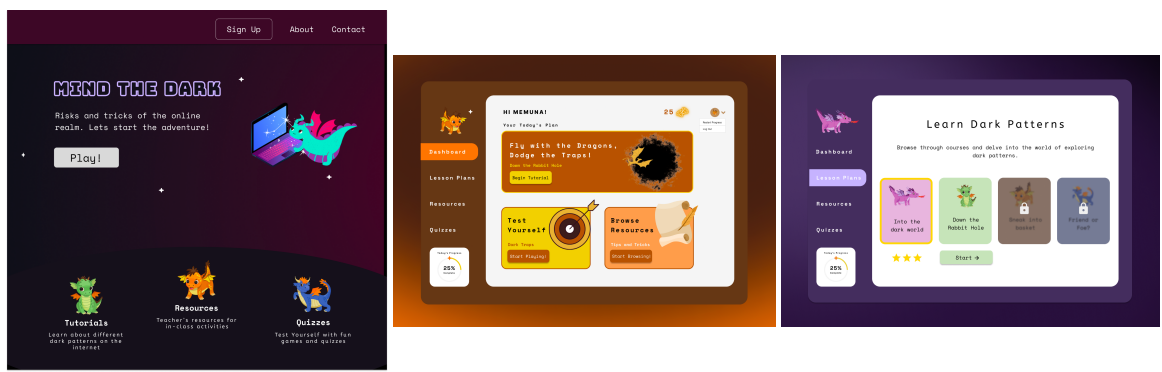

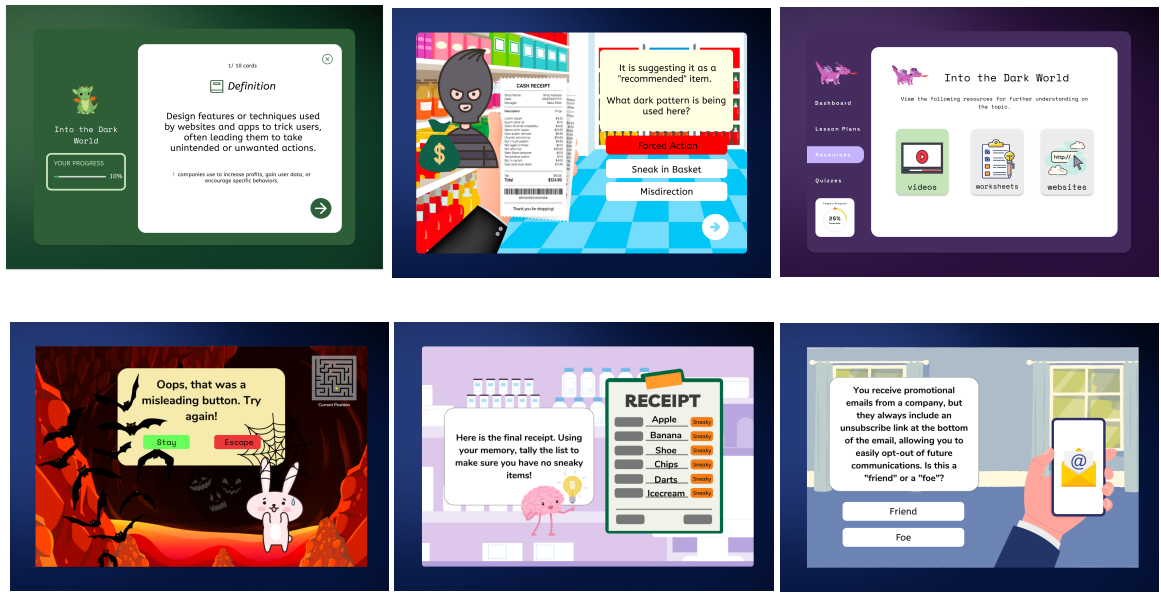

Fig. 2. App Landing Page, Dashboard, Homepage, Lesson Plan 1, Quiz 1, Resources, Quiz 2, Quiz 3, Quiz 4

The first module, "Into the Dark World" introduces dark patterns, providing an overview of their existence and prevalence in the online space and how deceptive techniques are used to manipulate their behaviors and actions. "Rabbit Hole" focuses on addictive dark patterns found in social media and video streaming platforms like YouTube and TikTok and how they can lead users down a rabbit hole of addictive content consumption. "Sneak Into Basket" delves into the dark patterns encountered while shopping online, mainly related to in-game purchases. It exposes children to the tactics employed to entice impulsive buying decisions and provides strategies to resist them while making informed choices. Lastly, "Friend or Foe" aims to highlight the importance of discerning trustworthy online sources and avoiding potential scams. Children learn to identify deceptive elements and distinguish between genuine and deceptive information, equipping them to navigate the online world safely.

## 6 EVALUATION

To evaluate our design, we conducted usability tests with 36 students from grades 3, 4 and 5 i.e. primary school children, who were different from the ones from our user research. We conducted the tests in groups of 5 to 6 children, and asked the participants to think aloud, while we observed how well they were able to understand the different dark patterns, and if they were able to successfully complete a lesson plan and its subsequent quiz. Once we concluded our tests, we conducted a short interview and asked participants to write their feedback about their experience with the app, including the vocabulary, animations, colors, features, and usability on pieces of papers as seen in figure 3. The purpose of conducting the testing in groups rather than with individuals was to simulate how it would be used in real life as the goal was to have the app be made part of the children's school curriculum.

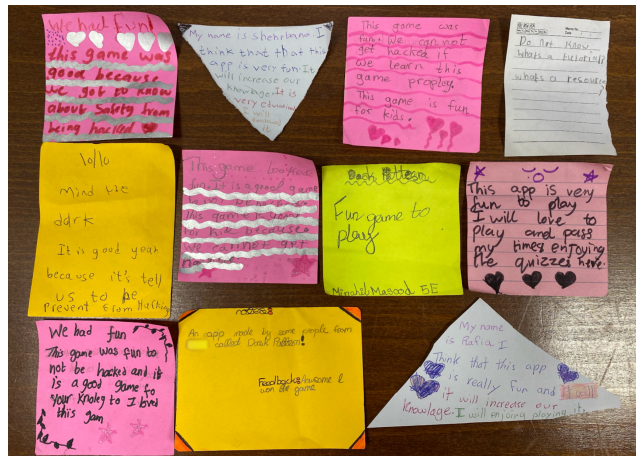
Fig. 3. Written feedback about our app from a few of the participants

The students went through each of the 3 lesson plans and their subsequent quizzes. When going through the lesson plans, most of the students didn't fully understand the lesson concept right after reading the definitions. However after going through the examples they began to show more understanding of the concept as they discussed it with each other and some students gave more examples from their personal experiences. For instance, one of the examples in lesson plan 1 was about the 'sneak into basket' dark pattern. In their discussion, the students mentioned their parents' or their own experiences with the deceptive design element which helped some of the other students understand it better. One student stated:

> "Oh this basket one is like that one time my mom was buying a shirt online and the store added matching pants automatically to the cart and she had to remove them" [P1, G3].

Another student had a similar comment:

> "Some websites also add some donation amount when you're about to pay for your stuff so you have to either pay it or cancel it which makes you feel guilty. But I think that's wrong because it should be our choice if we want to add it or not in the first place! Asking after adding is not good" [P8, G5]

After going through the examples, the reflection section further encouraged the students to engage in discussion by asking them to reflect upon what they had learned and think about any dark pattern they may have encountered in their lives, how it made them feel - did they feel like they had control over their actions? Some students stated that they felt angry and irritated about the additional clicks and the "mistaken clicks" (as they mentioned it - clicking on something by mistake when trying to get rid of something in a hurry) while some felt sad and guilty about the emotional emojis some websites used when urging them to perform a certain action. Towards the end of each lesson plan, the students had sufficient knowledge of the particular dark pattern in order to test it through the quiz. We noted the following statements from the students near the end of their discussions with their peers: "Oh now I understand it", "Ok this makes more sense now", "Alright now I'm ready for the quiz game".

Overall, the children found the game enjoyable to play which demonstrates our success in designing an intervention that children would want to play of their own accord. Children also commented on the utility of the game, mentioning that they liked how informative it was. Their validation of the usefulness of the game is not only indicative of the success of our goal but it also highlights the awareness of children. Children knew that dark patterns were an ever present issue in their technology use, they just had no words to describe them and thus did not know where to begin to ask for help. As soon as children began playing the game, they were able to recall deceptive encounters they experienced online. As a whole, the students acknowledged that the game was an entertaining and enjoyable way to learn about dark patterns, and something that they looked forward to using. The game increased students' knowledge about deceptive design and was conducive to their development of building self-reliance when facing unwanted situations online. Students thought the idea of our game, i.e. deceptive design, was very important and informational. A student stated:

> "I like the topic very much. It's very good and important because more and more children are using technology a lot nowadays so scams and hacking is very common" [P28, G5].

The game also encouraged discussion which helped students learn in a more collectivistic and wholesome manner. It inspired students to recount their personal experiences with dark patterns and share it with others for a better group-learning experience. While clicking on an incorrect option in the 'Down the Rabbit Hole' quiz simply because it was in the color green, a student recalled her personal experience with deceptive design:

> "Oh this happens a lot on random websites and the 'buy' or 'download' pop-up option is in green. My mind just thinks that green means yes so I click on it without reading it. Then I always go back and cross the pop-up" [P15, G4].

Another participant recalled his experience with dark patterns while going through one of the online shopping examples in a lesson plan:

> "I almost got scammed once in Roblox into buying more robux [the money in the game] than I wanted to. It added more in the cart by itself so I had to remove them" [P19, G4].

A majority of the participants thought the animations were very cute and some of them even asked when the app will be available for them to play on AppStores. A participant mentioned:

> "The rabbit is very cute and I love the colors [of the app]" [P6, G3].

Another student also had a similar comment:

> "I love the rabbit and the brain [another animated character]" [P7, G3].

The positive response of the children towards the type of animation and colors chosen for the game, evidences our design choices' success in engaging young children. While the majority of participants appreciated the design, animations and concept of the app, most of the students from grades 3 and 4 had difficulty understanding words like 'resources', 'techniques' and 'unintended'. Students from grade 5 however, had no trouble understanding the vocabulary in the lesson plans. This indicates that there is still a gap in the understanding of words between the youngest and oldest primary school children. This entails a revision of the choice of words in the design. Most participants were able to complete the lesson plans easily however whilst attempting the quizzes, participants did get stuck once or twice and selected the incorrect option.

    The teachers also had a very positive attitude towards our app. They praised our efforts in creating a useful intervention. The home room teacher of grade 4 stated:

> "It's great that you've come up with something so useful and necessary in this day and age, and rather than having kids play silly games, we should have them play games which are actually meaningful" [T1, G4].

All the home room teachers liked the idea of teaching children about deceptive design especially since they had started using technology at an early age. They suggested that such an informational game should be made part of their ICT syllabus as a class activity. The home room teacher of grade 5 mentioned:

> "This game could definitely be useful in educating kids on how to be more careful when using technology, especially since most of the younger ones use their parents' devices. My own son once accidently spent 120 dollars on Roblox because he uses my laptop which has my card info on it! So I had to email them to reverse the transaction" [T2, G5].

## 7 DISCUSSION

Our work studies children at the age at which they are transitioning from one stage of cognitive development to the next which is inline with the ages where they start gaining independent access to technology [5,6]. This is in contrast to prior research, which has not directly studied young children's interactions with dark patterns, in fact, studies were limited to studying adolescents and young teens [8, 10]. As first-time users exploring technology on their own, it is imperative that we form an understanding of the mental model of children, especially with regards to how they deal with manipulative design. With our research we have attempted to understand how children interact and engage with deceptive elements in technology and how they would react to it. Our research highlights an important aspect of child-computer interaction which is that children are cognisant of the danger of technology. We discovered that children were able to differentiate between "good" and "bad" interactions. This base understanding of children to be able to categorize helpful experiences from harmful ones was fundamental in the design process of our game. Some children, mostly the older ones, knew how to practice internet safety as a result of conversations with their parents; they already exercised caution when encountering manipulative design.

On the contrary, most children, mostly the younger ones, only had a vague understanding of misleading elements in technology and as a result weren't as vigilant as their other peers when it came to safely using technology. The content of our application assumes that the user has no prior knowledge of internet safety, enabling us to reinforce the existing knowledge of the first category of students while enlightening the second category of students. Overall, however, what both categories of children had in common was a lack of knowledge of the formal terminology used to describe and identify deceptive design. Thus, our primary focus of our game was to equip children with the right words to describe the "bad" interactions that they might encounter. By expanding their vocabulary, children would be better able to understand why someone would want to use certain dark patterns in certain situations and how to preemptively avoid them. Since children's use of technology will continue to increase as they age, our design and the awareness it provides will be able to educate and prepare them to safely navigate through it.

Our user research highlighted that when some children faced an unpleasant situation online they knew to turn to a guardian for help, as evidenced by the fact that many children would inform their parents if they received strange messages online. While this discovery highlights the astuteness of children in knowing when to ask for help, the goal of our intervention was to teach children self-reliance. Dealing with dark patterns is often a time-sensitive issue, which requires immediate action. Children who did turn to their guardians for help only did so after the negative interaction occurred e.g. after their account got hacked. Through our application, we want to guide children about how to preemptively avoid these situations and safely navigate technology in a way that if an unwanted situation were to present itself, they would be adept enough to resolve it on their own rather than solely relying on a guardian. As stated above, we wish to develop children's mental models in order to improve not only their tech literacy, but also their self-discipline and self-reliance.

Moreover, an important aspect of our intervention is collaborative learning as well as social learning. Through our study we observed that students engaged in discussions regarding dark patterns and shared their personal experiences with each other, which enhanced their overall learning and also encouraged students, who were otherwise shy, to raise their concerns and experiences regarding deceptive designs, thus taking part in the conversations. After the discussions, students were able to successfully complete the quizzes which showed knowledge retention. This is also in line with our literature, with regards to social and collaborative learning techniques enhancing critical thinking and overall understanding and retention of educational content [22, 23, 24].

Finally, it is important to note that our design is not meant to 'fix the problem' i.e. completely solve it, or teach children about specific dark patterns but rather to give awareness about such deceptive design elements which are ever evolving. So, the point of our design is to build children's critical thinking and navigation of online spaces so they're more cognizant of companies using legal methods to trick them and that they should be more careful about what they click on! Moreover, it must be pointed out that as researchers and designers, it is not the user's sole responsibility to protect themselves from deceptive UI but rather the onus of consumer protection and digital rights falls on the government. Since our research audience is from Pakistan, Pakistani law falls behind the international standard with no proper constitutional and statutory provisions on privacy [37]. Given its circumstances, Pakistani policy has not yet reached the stage where it considers deceptive design elements, and the impact it might have on its most vulnerable users, a pressing matter [37]. Thus, we have created our educational effort to give children the necessary tools and resources so that they can protect themselves.

## 8   LIMITATIONS AND FUTURE WORK

Our evaluation requires a long term longitudinal study in order to truly gauge the effectiveness of our educational intervention which was outside the scope of our time constraints. Thus our testing was limited to understanding children's initial insights into our game. While our evaluation showed that they absorbed new information about dark patterns and were able to relate it to past experiences, it did not show how well the children internalized this information and put it to use. Thus, we will conduct a semester-long evaluation process in the future to properly gauge the efficacy of our game, particularly the community learning aspect. We will also request the school to incorporate our game in their ICT syllabus for grades 3-5 and provide feedback after having the students use it for a term.

Furthermore, our game was designed keeping in mind that manipulative design is an ever present and dynamic landscape thus, we intentionally made the game scalable, so that additional modules about newer dark patterns can be included in the future. We have accounted for the fact that dark patterns will become more advanced and more undetectable in the future. This also necessitates that children be given a formal education about dark patterns, which is why we designed our application to be used at a school level. Our future work includes aiming to introduce our application into school curricula as a necessary part of teaching internet safety.

With the rampant spread of undetectable AI content and misinformation, the umbrella of internet safety topics to teach children is constantly widening. The framework with which we designed Mind The Dark proved to be successful in engaging and educating children about deceptive design. Thus, the same design guidelines can be used to design similar interventions to teach children about other emerging internet safety topics. Additionally, our principle of using less text and more visuals, minimizes the amount of information that is lost in translation and allows our application to be easily translatable to other languages. This ensures the universality of our design guidelines, allowing them to be applied at a global scale, in an effort to standardize education about internet safety.

## 9   SELECTION AND PARTICIPATION OF CHILDREN

For our user research, we visited a local private school and sought permission from the school principal to conduct focus groups with their primary and middle school students. We explained our topic and shared our research protocol including questions and lecture slides, so they could screen them. Once they granted us permission, we conducted focus groups within a section of each grade (3 to 7) in the presence of their home room teachers. Before beginning each focus group, we introduced ourselves to the students and explained the purpose of our research and how we would be conducting the focus groups so that children knew what to expect. We also took permission for audio-recording as well as note-taking from each home room teacher as well as the students and asked them to let us know if at any point they wanted to withdraw their participation from the study. We also mentioned that any direct quotes mentioned in our study will be anonymized. We also specified that the children were not required to answer any question if they didn't want to and that participation was voluntary. The children's personal identifiable information like their names, addresses etc. were not asked nor recorded. Moreover, we iterated that any data collected for this study would only be used for this research and not for any other purposes.